# Algebraic Derivation of the Schwarzschild Time Dilation Function

Andrew Theyken Bench[1]

Department of Physics and Astronomy, Franklin and Marshall College, 501 Harrisburg Avenue, Lancaster, Pennsylvania 17604

Using only a thought experiment and Einstein's correspondence principal, a model is derived that correctly predicts the Schwarzschild time dilation expression in limiting cases. The method requires almost no prerequisite knowledge from the student and is carried out with only algebraic techniques, allowing the introduction of a mathematical example of general relativity to introductory physics students.

01.30.la, 01.30.lb, 04.00.00

---

[1] andrew.bench@fandm.edu

**Introduction**
There has recently been a push in the literature[1] and in textbooks[2] to integrate the concepts of general relativity into the physics curriculum of undergraduate physics students. Such a push is certainly understandable considering the current state of theoretical physics of which "general relativity (GR) has become an integral and indispensable part."[3] It is somewhat, unfortunate, therefore, that introductory physics students, both at the undergraduate and high school level are rarely, if ever, presented with a mathematically motivated discussion of general relativity.

Opposition to such an attempt is certainly appropriate. Attempting to make any progress, in a mathematical sense, against GR, at a level appropriate to introductory students is difficult, but not impossible, if we motivate the discussion by attempting to "inform" the reader, rather than describe physical fact. In the derivation that follows, we will attempt to extend a general conceptual model of General Relativity into the realm of mathematical formalism.[4]

It is important to stress that what follows is an 'intuitive' approximation of general relativity; yet, what the model lacks in precision; it gains back in approachability. Prior to attempting this "general relativistic" problem, it would benefit the student immensely to see the derivation of time dilation in special relativity.[5] The derivation, herein given, is implicitly based on the model of special relativistic time dilation. Also, it would be beneficial to give a conceptual explanation of the correspondence principle[6] as the reference frame of this derivation is the ubiquitous accelerating *gedanke* spaceship.

**Method**
Imagine that we are floating in an interstellar spaceship, far away from any source of gravity. Also, we have just spoken to the captain, and he assures us that the ship is in an inertial frame, that is to say, all the force sensors onboard indicate there is no acceleration of any kind influencing the ship. We are very pleased by this because we want to conduct an experiment to determine how much time it takes a beam of light to go from the floor of the ship to the ceiling, a height $\Delta y$.

Being good theoreticians, we decide to work out how long it should take the beam of light to reach the ceiling with no acceleration:

$$\Delta y = \vec{c} \Delta t \quad (1)^7$$

$$\Delta t_i = \frac{\Delta y}{\vec{c}} \quad (1a)$$

$\Delta t_i \rightarrow$ time for light to travel a distance in inertial frame

We are ready to get underway to conduct the experiment when we get a message from the captain of the ship. He sends his apologies, but says we just received a distress signal and have to begin accelerating at a constant acceleration of $4.5 \times 10^{15} \frac{m}{s^2}$ ($a$) (in the +y direction). Undeterred, we decide simply to redo the calculation for (1a) in an accelerating reference frame.

Thus, we need to find an equation that will relate how much time it will take the light to reach the ceiling, in the presence of a constant acceleration. This is a little bit tricky to conceive conceptually, but let's change the model from a spaceship to standing on the earth in the presence of a gravitational force. Also, instead of shooting light upwards, let's throw a ball upwards and ask, in the presence of gravity, how long does it take the ball to go $\Delta y$? We can make this 'conceptual' change because of the Einstein equivalence principal. Now we can conceive very easily that the form of the equation we need is just one of the standard kinematics equations:

$$\Delta y = \vec{v} \Delta t_a + \frac{1}{2} \vec{a} \Delta t_a^2 \quad (2)$$

Thus, we solve equation (2) for $\Delta t_a$ using the quadratic equation:

$$\Delta t_a = \frac{-\vec{v} + \sqrt{\vec{v}^2 + 2\vec{a}\Delta y}}{\vec{a}} \quad (3)$$

We chose this particular solution of the quadratic because it refers to the "first" time the ball reaches $\Delta y$ on the upward part of its journey. Next, we transition back to the spaceship, where the ball we are "throwing" corresponds to light, so $v \rightarrow c$.

$$\Delta t_a = \frac{-\vec{c} + \sqrt{\vec{c}^2 + 2\vec{a}\Delta y}}{\vec{a}} \quad (3a)$$

Next, we decide to plug in numerical values[8] for both equation (1a) and equation (3a) and find out this surprising result:

$$\Delta t_i \rightarrow 3.33 \times 10^{-9} \text{ s}$$

$$\Delta t_a \rightarrow 3.42 \times 10^{-9} \text{ s}$$

The time interval for the accelerated frame is longer. Surely, this is a mistake? We can conceive that it makes sense in the case of the spaceship, because the "ceiling" of the spaceship moves upward, thus making the light travel a longer distance. But, if the Einstein correspondence principle is correct, then this suggests that in the presence of gravity, time is also dilated! Incredulous, we go back to the computer and graph equation (1) and equation (2) with the relevant numerical constants.

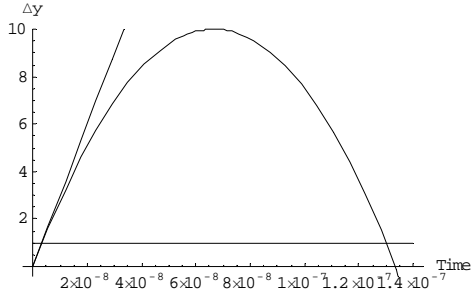

$\Delta y$ vs t. The horizontal time represents $\Delta y = 1$, the linear line represents the inertial model and the quadratic curve represents the accelerated model.

It seems like at every point, save zero, the linear curve for the inertial model is greater than the quadratic curve for the accelerated model. To be sure, we squeeze the time ordinal down very small:

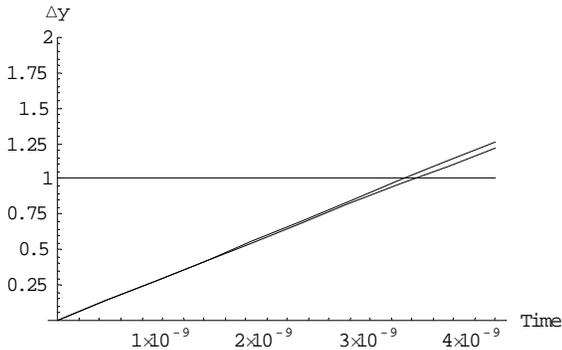

$\Delta y$ vs t. The horizontal time represents $\Delta y = 1$. The bottom curve is the accelerated model.

Incredibly, in the accelerated frame, or by the correspondence principal, the gravitationally affected frame, the time interval is longer. Time runs slower in the presence of gravity! Excited, we tell the captain and he asks us to try to relate the one time interval to the other.

We notice that equation (1a) and equation (2) and related to one another by the quantity $\Delta y$. Thus, we substitute equation (2) into equation (1a) and solve for $\Delta t_i$:

$$\Delta t_i = \frac{1}{\vec{c}}\left(\vec{c}\Delta t_a + \frac{1}{2}\vec{a}\Delta t_a^{\ 2}\right) \qquad (4)$$

$$\Delta t_i = \Delta t_a\left(1 + \frac{\vec{a}\Delta t_a}{2\vec{c}}\right) \qquad (5)$$

For reasons we will make clear momentarily, equation (5) is only accurate when $\Delta t_a$ is very small, in fact, when:

$$\Delta t_a \leq \frac{2\Delta y}{c} \qquad (6)$$

Equation (6) substituted into equation (5)

$$\Delta t_i \leq \Delta t_a\left(1 + \frac{\vec{a}\Delta y}{\vec{c}^{\ 2}}\right) \qquad (7)$$

Equation (7) represents the gravitational time dilation of our simple model.

**Comparison to Exact Solution**

We might ask ourselves how our model compares to actual theoretical results from GR. To do so we introduce a solution to the Einstein equations for a spherically symmetric mass, such as the Earth, known as the Schwarzschild solution.[9]

The exact solution to the time dilation problem in Schwarzschild space is given by the expression:

$$\Delta t_i = \Delta t_a\left(1 - \frac{2GM}{r\vec{c}^{\ 2}}\right)^{-1/2} \qquad (8)$$

To obtain a non-imaginary result, $\frac{2GM}{r\vec{c}^{\ 2}} \leq 1$; therefore, we can perform a binomial expansion of equation (8) in terms of $\frac{2GM}{r\vec{c}^{\ 2}}$. In almost all cases, we need only confine ourselves to the first term of the expansion:

$$\Delta t_i \approx \Delta t_a\left(1 + \frac{GM}{r\vec{c}^{\ 2}}\right) \qquad (9)$$

Furthermore, we can rewrite equation (9) in terms of the variables that we were using, corresponding to acceleration caused by gravity. To do so is a good exercise for the student. Also, we will make the change of variable, $r \rightarrow y$. We can do this because $y$ is simply one specific direction of a radial path.

$$\Delta t_i \approx \Delta t_a\left(1 + \frac{ay}{\vec{c}^{\ 2}}\right) \qquad (9a)$$

We compare equation (9a) to the model we derived, equation (7):

$$\Delta t_i \leq \Delta t_a \left(1 + \frac{a\Delta y}{\vec{c}^2}\right) \quad (7)$$

The correspondence between equation (7) and equation (9a) is really quite amazing. True, we had to use some mathematical 'trickery' to get the equations to look like one another, but the mathematical form of the equations is precisely the same (save a nonessential delta and an inequality sign). This suggests, at the very least, that the assumptions that we made in our derivation were not at all baseless, and that GR really does behave in a way consistent with our assumptions. Below, we will think carefully about the assumptions and approximations that we made.

**Error and Justification**

At equation (6) we noted that $\Delta t_a \leq \frac{2\Delta y}{\vec{c}}$. Why? The point is actually very important and very revealing. We used as our fundamental model for an accelerated frame equation (2). However, there was a subtle Newtonian assumption in equation (2) that we have not discussed until now—the concept of constant acceleration. Imagine that we accelerated a rocket constantly at $4.5 \times 10^{15} \frac{m}{s^2}$. If acceleration were constant, at $t = 6.66 \times 10^{-8}$ s, we would achieve the speed of light! A nanosecond thereafter we would surpass the speed of light. This is not allowed. Relativistically, there is no such thing as constant acceleration. Yet, this is what we used as our model. Thus, to prevent unwanted relativistic side effects of non-constant acceleration, we must only concern ourselves when the speed of our rocket is well below the speed of light, thus, very short times. Determining how short "very short" is takes some foresight, but can be determined as follows.

Recall that equation (3a) contains the quantity $\sqrt{\vec{c}^2 - 2\vec{a}\Delta y}$.[10] If $\vec{a}$ becomes incredibly large, then we get an imaginary result, which, physically, corresponds to the fact that the light beam never hits the ceiling, because the spaceship is now traveling faster than a light beam. We know this cannot be true because the space ship CANNOT go faster than the speed of light equation (9); therefore:

$$\vec{c}^2 - 2\vec{a}\Delta y \geq 0 \quad (8)$$

$$\vec{a} \leq \frac{\vec{c}^2}{2\Delta y} \quad (8a)$$

$$\vec{a}\Delta t_a \leq \vec{c} \quad (9)$$

Equation (8a) is substituted into equation (9)

$$\Delta t_a \leq \frac{2\Delta y}{\vec{c}} \quad (6)$$

Thus, we now understand the inequality in equation (7). Under limiting conditions, very short $\Delta t_a$, very small $\vec{a}$'s, or very small $\Delta y$'s, equation (7) $\approx$ equation (9a). If any of those factors increase, we move away from our assumed constant acceleration and the model begins to diverge from the Schwarzschild model.

**Discussion**

Using just the precept of a thought experiment and the notion of Einsteinian correspondence, we have determined a simple, but effective, model to explain gravitational time dilation. Einstein would have been proud! Not only do the results make conceptual sense, but in the limit, they become the solution of the exact Schwarzschild spacetime. What is more, we developed all of the concepts with very simple algebra, at a mathematical level appropriate to introductory college students and advanced high school students. If the instructor wished to extend the results herein presented she could certainly consider the situation in terms of the relativistic acceleration function.[11] Such a derivation would certainly demand much more mathematical sophistication[12] from the student, but would require far fewer assumptions, and may, in that respect, actually be clearer to a mathematically prepared student.

---